\documentclass[onecolumn]{revtex4}
\usepackage{ragged2e}
\usepackage{amsmath}
\usepackage{graphicx}
\usepackage{bm}
\usepackage{tabularx} 
\usepackage{array} 

\usepackage[usenames,dvipsnames]{color}

\definecolor{darkblue}{RGB}{0,0,196}

\begin{document}

\title{A hybrid model for multi-particle production and
multi-fragment emission in electron-nucleus collisions at the
forthcoming Electron-Ion Collider}\vspace{0.5cm}

\author{Ting-Ting~Duan$^{1,}${\footnote{202312602001@email.sxu.edu.cn}},
Sahanaa~B$\rm\ddot{u}$riechin$^{1,}${\footnote{202201101236@email.sxu.edu.cn}},
Hai-Ling~Lao$^{2,}${\footnote{hailinglao@163.com;
hailinglao@pku.edu.cn}},
Fu-Hu~Liu$^{1,}${\footnote{Correspondence: fuhuliu@163.com;
fuhuliu@sxu.edu.cn}},
Khusniddin~K.~Olimov$^{3,4,}${\footnote{Correspondence:
khkolimov@gmail.com; kh.olimov@uzsci.net}}}

\affiliation{$^1$Institute of Theoretical Physics, State Key
Laboratory of Quantum Optics Technologies and Devices \& \\
Collaborative Innovation Center of Extreme Optics, Shanxi
University, Taiyuan 030006, China
\\
$^2$Department of Science Teaching, Beijing Vocational College of
Agriculture, Beijing 102442, China
\\
$^3$Laboratory of High Energy Physics, Physical-Technical
Institute of Uzbekistan Academy of Sciences, Chingiz Aytmatov Str.
2b, Tashkent 100084, Uzbekistan
\\
$^4$Department of Natural Sciences, National University of Science
and Technology MISIS (NUST MISIS), Almalyk Branch, Almalyk 110105,
Uzbekistan}

\begin{abstract}

\vspace{0.5cm}

\noindent {\bf Abstract:} To present a prediction of the
multi-particle production and multi-fragment emission in
electron-nucleus ($eA$) collisions at the forthcoming Electron-Ion
Collider (EIC), a simple hybrid model which is based on the
multi-source thermal model and the ideal gas model is proposed in
this article. According to the hybrid model, some statistical laws
such as the two-component Erlang distribution and others are
presented, which means a two-source production. These statistical
laws are hopeful to describe the bulk properties of multiple
particles produced in the scattering of electron-nucleon ($eN$)
and multiple fragments emitted in the fragmentation of excited
residual nucleus. Although both the scattering and fragmentation
can occur in $eA$ collisions at the EIC, their two-sources are
different. In $eN$ scattering, the multiple particles come from
the soft excitation and hard scattering processes respectively,
which are classified as two different types of events. In nuclear
fragmentation, the multiple fragments come from the cold and hot
sources which exist in the same excited nucleus.
\\
\end{abstract}

\maketitle

\parindent=15pt

\section{Introduction}

Numerous particles and fragments have been measured in high-energy
nuclear collision experiments~\cite{1,2,3}. It is understood that
these multiple particles and fragments stem from different sources
with varying production mechanisms~\cite{4,5,6,7,8}. Multiple
particles may be produced within the region occupied by
participants; however, only a few may emerge through cascade
processes occurring in spectator regions when available.
Similarly, while numerous fragments can be emitted from
spectators' regions---only a limited number of light fragments
might be released from participant areas.

At the nucleon level, participants and spectators consist of
nucleons along with their clusters~\cite{9,10,11,12,13}. At the
partonic level---participants include quarks and gluons while
spectators also comprise similar
constituents~\cite{14,15,16,17,18}. In nuclear collisions induced
by hadrons---the hadron acts as a participant that is
(approximately) equivalent to a nucleon. In nuclear collisions
induced by a lepton, the lepton acts as a participant that is
approximately equivalent to a parton. The statistical distribution
laws governing the multiple particles and fragments produced in
these interactions are of particular interest to us.

Generally, both types of products exhibit similarities and
differences in their statistical distribution
laws~\cite{19,20,21,22,23,24,25,26}. Our investigations reveal
that the multiplicity and transverse momentum distributions of
multiple particles are analogous; they adhere to the two-component
Erlang distribution within the framework of a multi-source thermal
model~\cite{27,28}. Similarly, the multiplicity distributions for
both multiple particles and fragments follow this same
distribution and model. However, it is important to note that the
emission sources for multiple particles differ from those for
fragments. The study of multiple particles can be conducted at the
parton level while investigations into multiple fragments can be
performed at the nucleon level.

Understanding the statistical distribution laws followed by
multiple particles and fragments generated in electron-nucleus
($eA$) collisions at the forthcoming Electron Ion Collider (EIC)
holds significant importance. Firstly, $eA$ collisions represent a
deep inelastic scattering process between an incoming projectile
electron ($e$) and a target nucleon ($N$) within the target
nucleus, where $N$ may be either a proton ($p$) or neutron ($n$).
Secondly, $eN$ scattering constitutes an even deeper inelastic
scattering process involving interactions between incoming
electron $e$ with target quarks ($q$) and/or gluons ($g$) present
within nucleon $N$. Consequently, $eA$ collisions encompass both
an $eN$ scattering event as well as an excited residual nucleus.
In contrast, $eN$ scattering involves one or more instances of
$eq$ (or $eg)$ scattering.

Compared to nuclear collisions induced by hadrons or nuclei, $eA$
collisions offer a cleaner environment without the complexity of
background products, serving as an essential baseline for both
hadron-nucleus and nucleus-nucleus collisions. The bulk properties
of the final-state particles and fragments generated in $eA$
collisions also establish a fundamental reference point, crucial
for investigating the unique characteristics of new particle
production. Furthermore, when juxtaposed with $ee$ collisions, the
nuclear effects observed in $eA$ collisions can significantly
influence experimental outcomes, making the study of these effects
an intriguing area of research.

In this article, we introduce a straightforward hybrid model for
multi-particle production and multi-fragment emission. We present
statistical distribution laws comprising two components, which
anticipate the bulk properties of the final-state particles and
fragments produced in $eA$ collisions at the EIC.

\section{The hybrid model for $eA$ collisions at the EIC}

It is widely accepted that $eA$ collisions at the EIC represent a
deep inelastic scattering process at the nucleon level. However,
it is important to note that the actual interaction occurring in
$eA$ collisions can be understood as $eN$ collisions, which also
constitute a deep inelastic scattering process but occur at the
parton level. By excluding the participating nucleon involved in
the deep inelastic scattering within $eA$ collisions, one can
obtain the remnant portion of the nucleus. This remnant is
referred to as a spectator and has the potential to form an
excited nucleus that subsequently fragments into various
components.

\subsection{Multi-particle production}

It is believed that $eN$ scattering during $eA$ collisions leads
to multi-particle production process. Several thermal and
statistical models related to multi-particle production may be
applied to analyze this phenomenon within $eN$ scattering. For
instance, one might consider modelling the $eN$ scattering process
using a multi-source thermal model~\cite{27,28}, wherein each
nucleon is treated as an extensive system composed of valence
quarks, sea quarks, and gluons. Given that $eN$ scattering can
proceed through both soft and hard processes, event samples are
typically mixtures of soft excitations and hard scatterings. In
this context, an electron interacts with several ($m_1-1$) sea
quarks and/or gluons during the soft excitation phase---denoted as
Process One---while another scenario involves an electron
interacting with a valence quark during hard scattering---referred
to as Process Two.

Regarding multiplicity ($n_{ch}$) distribution of charged
particles, each participant (whether electron or parton) involved
in the soft process contributes a quantity denoted by $n_{i1}$,
which follows an exponential distribution $f_{n_{i1}}(n_{i1})$
characterized by its parameter $\langle n_{i1} \rangle$,
representing its average value. Similarly, each participant
engaged in the hard process contributes a quantity represented by
$n_{i2}$, adhering to another exponential distribution
$f_{n_{i2}}(n_{i2})$, but defined by its own parameter $\langle
n_{i2} \rangle$. Each participant is regarded as having an
effective energy source~\cite{22,23,24}. The behavior of multiple
participants is hopeful to be described by the multi-source
thermal model~\cite{27,28}.

One has a general exponential distribution
\begin{align}
f_{n_{i}}(n_{i})=\frac{1}{\langle n_{i}\rangle}
\exp\left(-\frac{n_{i}}{\langle n_{i}\rangle}\right),
\end{align}
which is normalized to 1. Here, a subscript $n_i$ is used to
distinguish the distribution from others which will be discussed
later. The Erlang distribution is
\begin{align}
f_{n_{ch},E}(n_{ch})=\frac{n_{ch}^{m_j-1}}{(m_j-1)!\langle
n_{i}\rangle^{m_j}} \exp\left(-\frac{n_{ch}}{\langle n_{i}\rangle}
\right),
\end{align}
which is the fold of $m_j$ ($j=1$ or 2) exponential distributions,
where $n_i=n_{i1}$ ($n_i=n_{i2}$) and $\langle
n_{i}\rangle=\langle n_{i1}\rangle$ ($\langle n_{i}\rangle=\langle
n_{i2}\rangle$) are for the soft (hard) process. The numbers of
participants in the soft and hard processes are $m_1$ and $m_2$
($=2$) respectively. Eq. (2) is normalized to 1 and results in the
average of $n_{ch}$ to be
\begin{align}
\langle n_{ch}\rangle=\int n_{ch}
f_{n_{ch},E}(n_{ch})dn_{ch}=m_j\langle n_i\rangle.
\end{align}
The $n_{ch}$ distribution of charged particles measured in the
final state is a two-component Erlang distribution (the
superposition of two Erlang distributions)~\cite{27} given by
\begin{align}
f_{n_{ch},2E}(n_{ch})=\frac{k_1 n_{ch}^{m_1-1}}{(m_1-1)!\langle
n_{i1}\rangle^{m_1}} \exp\left(-\frac{n_{ch}}{\langle
n_{i1}\rangle} \right) +\frac{(1-k_1) n_{ch}}{\langle
n_{i2}\rangle^2} \exp\left(-\frac{n_{ch}}{\langle n_{i2}\rangle}
\right),
\end{align}
where $k_1$ and $1-k_1$ are the contribution fractions of the soft
and hard processes. Eq. (4) is normalized to 1 and gives
\begin{align}
\langle n_{ch}\rangle=\int n_{ch}
f_{n_{ch},2E}(n_{ch})dn_{ch}=k_1m_1\langle
n_{i1}\rangle+2(1-k_1)\langle n_{i2}\rangle.
\end{align}

The transverse momentum ($p_T$) distribution of charged particles
can be fitted using several functions, including the standard
distributions (Fermi--Dirac/Bose--Einstein or Maxwell--Boltzmann)
and their multi-component forms, the Tsallis distribution and its
multi-component variants, inverse power laws and their alternative
representations, as well as various superpositions of these
functions. Additionally, some Monte Carlo event generators have
been employed for this purpose~\cite{29,30}. Within the framework
of a multi-source thermal model, the two-component Erlang
distribution has been utilized to fit the $p_T$ distribution of
charged particles~\cite{28}.

Similar to the distributions of $n_i$ and $n_{ch}$, both the
transverse momentum $p_{ti}$ distribution contributed by each
participant and the overall $p_T$ distribution exhibit analogous
expressions:
\begin{align}
f_{p_{ti}}(p_{ti})=\frac{1}{\langle p_{ti}\rangle}
\exp\left(-\frac{p_{ti}}{\langle p_{ti}\rangle}\right),
\end{align}
\begin{align}
f_{p_T,E}(p_{T})=\frac{p_T^{M_j-1}}{(M_j-1)!\langle
p_{ti}\rangle^{M_j}} \exp\left(-\frac{p_T}{\langle p_{ti}\rangle}
\right),
\end{align}
\begin{align}
f_{p_{T},2E}(p_{T})=\frac{K_1 p_{T}^{M_1-1}}{(M_1-1)!\langle
p_{ti1}\rangle^{M_1}} \exp\left(-\frac{p_{T}}{\langle
p_{ti1}\rangle} \right)+\frac{(1-K_1) p_{T}}{\langle
p_{ti2}\rangle^2} \exp\left(-\frac{p_{T}}{\langle p_{ti2}\rangle}
\right).
\end{align}
Here, $M_j$ ($j=1$ or 2) is the number of participant partons in
the $j$-th process (soft or hard process) with $M_2=2$ in general,
$K_1$ ($1-K_1$) is the contribution fraction of the soft (hard)
process, and $\langle p_{ti1}\rangle$ ($\langle p_{ti2}\rangle$)
is the average contribution of each participant in the soft (hard)
process. Eqs. (7) and (8) are normalized to 1 and give the
averages of $p_T$ to be
\begin{align}
\langle p_T\rangle=\int p_{ti}f_{p_T,E}(p_{ti})dp_{ti}=M_j\langle
p_{ti}\rangle
\end{align}
and
\begin{align}
\langle p_T\rangle=\int p_{ti}f_{p_T,2E}(p_{ti})dp_{ti}
=K_1M_1\langle p_{ti1}\rangle+2(1-K_1)\langle p_{ti2}\rangle,
\end{align}
respectively. According to a thermal-related
method~\cite{30a,30b}, one has the temperature $T=\langle
p_T\rangle/3.07$.

Eqs. (1) and (6), (2) and (7), along with (4) and (8), share
similar forms but differ in terms of variables and parameters. In
Eq. (4), the free parameters are denoted as $k_1$, $m_1$, $\langle
n_{i1}\rangle$, and $\langle n_{i2}\rangle$. Conversely, in Eq.
(8), they are represented by $K_1$, $M_1$, $\langle
p_{ti1}\rangle$, and $\langle p_{ti2}\rangle$. The values assigned
to $k_1$ and $K_1$ ($m_1$ and $M_1$) in Eqs. (4) and (8) may vary
due to differences in event samples.

According to the two-cylinder model~\cite{30c,30d}, which is a
component of the multi-source thermal model~\cite{27,28}, whether
considering soft excitation or hard scattering process, both
incoming projectile $e$ and target $N$ can penetrate one another.
This interaction leads to the formation of a projectile string
alongside a target string. In rapidity space, these strings
distribute uniformly across their respective rapidity ranges:
[$y_{P\min}, y_{P\max}$] for projectiles; [$y_{T\min}, y_{T\max}$]
for targets. Notably, while leading target protons occupy rapidity
position at $y_T$, it is important to emphasize that neither
projectile $e$---which does not serve as a leading particle---nor
target neutrons should be classified among leading charged
particles.

When the strings are broken, for an isotropic emission source
located at the rapidity $y_x$, the pseudorapidity distribution of
charged particles is given by
\begin{align}
f_{\eta}(\eta,y_x)=\frac{1}{2\cosh^2(\eta-y_x)}.
\end{align}
In experiments, the pseudorapidity distribution of charged
particles is given by
\begin{align}
f_{\eta}(\eta)=&\,\frac{K_P}{2(y_{P\max}-y_{P\min})}
\int_{y_{P\min}}^{y_{P\max}}\frac{dy_x}{\cosh^2(\eta-y_x)}\nonumber\\
&+\frac{K_T}{2(y_{T\max}-y_{T\min})}
\int_{y_{T\min}}^{y_{T\max}}\frac{dy_x}{\cosh^2(\eta-y_x)} +
\frac{1-K_P-K_T}{2\cosh^2(\eta-y_T)},
\end{align}
where $K_P$, $K_T$, and $1-K_P-K_T$ are the contribution fractions
of the projectile string, target string, and leading target
protons, respectively.

The length, $L_{y_P}=y_{P\max}-y_{P\min}$
($L_{y_T}=y_{T\max}-y_{T\min}$), of the projectile (target) string
in the soft process is longer than that in the hard process.
Because of the contribution fraction of the soft process being
larger, the average string length in the soft and hard processes
is predominantly determined by the soft process. The kinematic
relationship between projectile and target strings is
quantitatively expressed through the rapidity gap parameter:
$\Delta y=y_{P\min}-y_{T\max}$. This dimensionless quantity
exhibits three characteristic regimes: $\Delta y>0$, $\Delta y=0$,
and $\Delta y<0$, which implies the spatially separated
configuration, marginally connected state, and geometrically
overlapping domain, respectively. Thus, the sign convention
establishes a bijective correspondence between algebraic values
and physical interpretations.

\subsection{Multi-fragment emission}

The decay or fragmentation of the spectator (the excited nucleus)
is characterized by a process of multi-fragment emission
originating from two distinct sources, each with its own
temperature~\cite{31}. These sources differ from the two-component
distributions typically observed in charged particles. The local
region associated with $eN$ scattering serves as the hot source,
exhibiting a high temperature, while another portion within the
spectator acts as a cold source with a lower temperature. Both
sources exist in their respective equilibrium states and contain
several nucleons that are treated as contributors.

Each contributor, denoted as the $i$-th element in either the cold
(hot) source, contributes an amount $N_{iL}$ ($N_{iH}$) to the
overall multiplicity $N_F$ of all nuclear fragments~\cite{32}. It
is assumed that $N_i$ ($N_{iL}$ or $N_{iH}$) follows an
exponential distribution:
\begin{align}
f_{N_{i}}(N_{i})=\frac{1}{\langle N_{i}\rangle}
\exp\left(-\frac{N_{i}}{\langle N_{i}\rangle}\right),
\end{align}
where the parameter $\langle N_{i}\rangle$ is the average of
$N_i$. The sum of the contributions of $m_j$ ($j=L$ or $H$)
contributors is an Erlang distribution, given by
\begin{align}
f_{N_{F},E}(N_{F})=\frac{N_{F}^{m_j-1}}{(m_j-1)!\langle
N_{i}\rangle^{m_j}} \exp\left(-\frac{N_{F}}{\langle N_{i}\rangle}
\right)
\end{align}
with the average of $N_F$ to be
\begin{align}
\langle N_{F}\rangle=\int N_{F}
f_{N_{F},E}(N_{F})dN_{F}=m_j\langle N_i\rangle.
\end{align}
The sum of the contributions of the cold and hot sources are a
two-component Erlang distribution, which is
\begin{align}
f_{N_{F},2E}(N_{F})=\frac{k_L N_{F}^{m_L-1}}{(m_L-1)!\langle
N_{iL}\rangle^{m_L}} \exp\left(-\frac{N_{F}}{\langle
N_{iL}\rangle} \right)
+\frac{(1-k_L)N_{F}^{m_H-1}}{(m_H-1)!\langle N_{iH}\rangle^{m_H}}
\exp\left(-\frac{N_{F}}{\langle N_{iH}\rangle} \right)
\end{align}
with
\begin{align}
\langle N_{F}\rangle=\int N_{F}
f_{N_{F},2E}(N_{F})dN_{F}=k_Lm_L\langle N_{iL}\rangle+
(1-k_L)m_H\langle N_{iH}\rangle.
\end{align}
Here, $k_L$ ($1-k_L$) denotes the contribution fraction of the
cold (hot) source.

In $eA$ collisions, we define the direction of the incoming
nucleus along the $Oz$ axis. Consequently, the reaction plane
corresponds to the $xOz$ plane. This establishes a
three-dimensional Cartesian coordinate system for analysis. A
general thermal and statistical model can be employed to describe
nuclear fragment emissions within the rest frame of each
considered source. Subsequently, one can derive both transverse
momentum ($p_T$) and angular ($\theta$) distributions of nuclear
fragments in either laboratory or center-of-mass reference
frames~\cite{33}.

Typically, the excitation degree of the spectator remains
relatively low; thus, temperatures during nuclear fragmentation
are on the order of several MeV or slightly higher. The classical
ideal gas model may be utilized to characterize both $p_T$ and
$\theta$ distributions for these nuclear fragments. In relation to
their respective rest frames---cold or hot sources---the
$x$-component momentum $p_x$, alongside $y$-component momentum
$p_y$, and $z$-component momentum $p_z$, adheres to Gaussian
distribution patterns:
\begin{align}
f_{p_{x,y,z}}(p_{x,y,z})=\frac{1}{\sqrt{2\pi}\sigma_{p}}
\exp\left(-\frac{p_{x,y,z}^2}{2\sigma_{p}^2} \right),
\end{align}
where $\sigma_{p}=\sigma_{pL}=\sqrt{m_FT_L}$
($\sigma_{p}=\sigma_{pH}=\sqrt{m_FT_H}$) is the width of momentum
distribution of nuclear fragments emitted from the cold (hot)
source, $T_L$ ($T_H$) is the temperature of the cold (hot) source,
and $m_F$ is the mass of the considered nuclear fragment.

Naturally, the transverse momentum $p_T=\sqrt{p_x^2+p_y^2}$ obeys
the Rayleigh distribution, given by
\begin{align}
f_{p_{T},R}(p_{T})=\frac{p_T}{\sigma_p^2}
\exp\left(-\frac{p_{T}^2}{2\sigma_p^2} \right)
\end{align}
with
\begin{align}
\langle p_T\rangle=\int p_T f_{p_{T},R}(p_{T})dp_T=\sigma_p
\sqrt{\frac{\pi}{2}}.
\end{align}
Considering the two-temperature case, one has
\begin{align}
f_{p_{T},2R}(p_{T})= \frac{K_Lp_T}{\sigma_{pL}^2}
\exp\left(-\frac{p_{T}^2}{2\sigma_{pL}^2} \right)
+\frac{(1-K_L)p_T}{\sigma_{pH}^2}
\exp\left(-\frac{p_{T}^2}{2\sigma_{pH}^2} \right)
\end{align}
with
\begin{align}
\langle p_T\rangle=\int p_T f_{p_{T},2R}(p_{T})dp_T=\big[
K_L\sigma_{pL}+(1-K_L)\sigma_{pH}\big] \sqrt{\frac{\pi}{2}}.
\end{align}
Here, $K_L$ ($1-K_L$) denotes the contribution fraction of the
cold (hot) source. The values of $k_L$ and $K_L$ in the
expressions of $N_F$ and $p_T$ distributions may be different due
to different event samples.

In laboratory or center-of-mass reference frames, it is
approximately true that per nucleon momentum conservation holds
for nuclear fragments relative to incoming nuclei; hence one finds
that $p_T=p \sin\theta\approx p\theta$, where $p$ represents the
momentum associated with any given nuclear fragment under
consideration. The angular distribution obeys approximately the
Rayleigh form:
\begin{align}
f_{\theta,R}(\theta)\approx\frac{\theta}{\sigma_{\theta}^2}
\exp\left(-\frac{\theta^2}{2\sigma_{\theta}^2} \right)
\end{align}
with
\begin{align}
\langle \theta\rangle=\int \theta f_{\theta,R}(\theta)
d\theta\approx \sigma_{\theta} \sqrt{\frac{\pi}{2}}.
\end{align}
Here $\sigma_{\theta}=\sigma_p/p=\sigma_{pL}/p=\sigma_{\theta L}$
($\sigma_{\theta}=\sigma_p/p=\sigma_{pH}/p=\sigma_{\theta H}$) is
the width of angular distribution of nuclear fragments emitted
from the cold (hot) source. Considering the two-temperature case,
one has
\begin{align}
f_{\theta,2R}(\theta)\approx \frac{K_L\theta}{\sigma_{\theta L}^2}
\exp\left(-\frac{\theta^2}{2\sigma_{\theta L}^2} \right) +
\frac{(1-K_L)\theta}{\sigma_{\theta H}^2}
\exp\left(-\frac{\theta^2}{2\sigma_{\theta H}^2} \right)
\end{align}
with
\begin{align}
\langle \theta\rangle=\int \theta f_{\theta,2R}(\theta)
d\theta\approx \big[K_L \sigma_{\theta L}+(1-K_L) \sigma_{\theta
H} \big]\sqrt{\frac{\pi}{2}}.
\end{align}

If we consider the relativistic ideal gas model~\cite{34,35}, Eqs.
(18), (19), and (21) can be written as the following
forms~\cite{36}
\begin{align}
f_{p_{x,y}}(p_{x,y})= C_{x,y}(T)\int_{-\infty}^{\infty}
\int_{-\infty}^{\infty}
\exp\left(-\frac{\sqrt{p_x^2+p_y^2+p_z^2+m_F^2}}{T}\right)dp_zdp_{y,x},\nonumber\\
f_{p_z}(p_z)=C_z(T)\int_0^{\infty}p_T
\exp\left(-\frac{\sqrt{p_T^2+p_z^2+m_F^2}}{T}\right)dp_T,\qquad\quad\,\,\,\,
\end{align}
\begin{align}
f_{p_T}(p_T)=C_T(T)p_T\int_{-\infty}^{\infty}
\exp\left(-\frac{\sqrt{p_T^2+p_z^2+m_F^2}}{T}\right)dp_z,
\end{align}
\begin{align}
f_{p_T}(p_T)=&\, K_LC_T(T_L)p_T\int_{-\infty}^{\infty}
\exp\left(-\frac{\sqrt{p_T^2+p_z^2+m_F^2}}{T_L}\right)dp_z
\nonumber\\
&+(1-K_L)C_T(T_H)p_T\int_{-\infty}^{\infty}
\exp\left(-\frac{\sqrt{p_T^2+p_z^2+m_F^2}}{T_H}\right)dp_z.
\end{align}
Here, $C_{x,y,z,T}(T)$ are the normalization related to the
temperature $T$, which result in
$\int_{-\infty}^{\infty}f_{p_{x,y,z,T}}(p_{x,y,z,T})dp_{x,y,z,T}=1$.
Correspondingly, Eqs. (23) and (25) are rewritten as
\begin{align}
f_{\theta}(\theta)\approx
C_{\theta}(T)\theta\int_{-\infty}^{\infty}
\exp\left(-\frac{\sqrt{p^2\theta^2+p_z^2+m_F^2}}{T}\right)dp_z,
\end{align}
\begin{align}
f_{\theta}(\theta)\approx &\,
K_LC_{\theta}(T_L)\theta\int_{-\infty}^{\infty}
\exp\left(-\frac{\sqrt{p^2\theta^2+p_z^2+m_F^2}}{T_L}\right)dp_z
\nonumber\\
&+(1-K_L)C_{\theta}(T_H)\theta\int_{-\infty}^{\infty}
\exp\left(-\frac{\sqrt{p^2\theta^2+p_z^2+m_F^2}}{T_H}\right)dp_z,
\end{align}
where $C_{\theta}(T)$ represents the normalization associated with
the temperature $T$, leading to the condition
$\int_0^{\pi}f_{\theta}(\theta)d\theta=1$. Deriving specific
arithmetic or functional expressions for $\langle p_T\rangle$ and
$\langle \theta\rangle$ from Eqs. (28)--(31) can be cumbersome;
however, a numerical method may be employed if necessary.

Alternatively, within the framework of the relativistic ideal gas
model~\cite{34,35}, in terms of rapidity $y$ and a constrained
rapidity range $[y_{\min},y_{\max}]$, Eqs. (10)--(14) can be
reformulated as indicated in~\cite{36},
\begin{align}
f_{p_{x,y}}(p_{x,y})=
C_{x,y}(T)\int_{-\infty}^{\infty}\sqrt{p_x^2+p_y^2+m_F^2}
\int_{y_{\min}}^{y_{\max}}\cosh y \times
\exp\left(-\frac{\sqrt{p_x^2+p_y^2+m_F^2}\cosh
y}{T}\right)dydp_{y,x},\nonumber
\end{align}
\begin{align}
f_{p_z}(p_z)=C_z(T)\int_0^{\infty}p_T
\exp\left(-\frac{\sqrt{p_T^2+p_z^2+m_F^2}}{T}\right)dp_T,
\end{align}
\begin{align}
f_{p_T}(p_T)=C_T(T)p_T\sqrt{p_T^2+m_F^2}\int_{y_{\min}}^{y_{\max}}
\cosh y \times \exp\left(-\frac{\sqrt{p_T^2+m_F^2}\cosh
y}{T}\right)dy,
\end{align}
\begin{align}
f_{p_T}(p_T)=&\, K_LC_T(T_L)p_T\sqrt{p_T^2+m_F^2}
\int_{y_{\min}}^{y_{\max}} \cosh y \times
\exp\left(-\frac{\sqrt{p_T^2+m_F^2}\cosh
y}{T_L}\right)dy \nonumber\\
&+(1-K_L)C_T(T_H)p_T\sqrt{p_T^2+m_F^2}\int_{y_{\min}}^{y_{\max}}
\cosh y \times \exp\left(-\frac{\sqrt{p_T^2+m_F^2}\cosh
y}{T_H}\right)dy,
\end{align}
\begin{align}
f_{\theta}(\theta)\approx C_{\theta}(T)\theta
\sqrt{p^2\theta^2+m_F^2}\int_{y_{\min}}^{y_{\max}} \cosh y \times
\exp\left(-\frac{\sqrt{p^2\theta^2+m_F^2}\cosh y}{T}\right)dy,
\end{align}
\begin{align}
f_{\theta}(\theta)\approx &\, K_L C_{\theta}(T_L)\theta
\sqrt{p^2\theta^2+m_F^2}\int_{y_{\min}}^{y_{\max}} \cosh y \times
\exp\left(-\frac{\sqrt{p^2\theta^2+m_F^2}\cosh y}{T_L}\right)dy
\nonumber\\
&+(1-K_L)C_{\theta}(T_H)\theta
\sqrt{p^2\theta^2+m_F^2}\int_{y_{\min}}^{y_{\max}} \cosh y \times
\exp\left(-\frac{\sqrt{p^2\theta^2+m_F^2}\cosh y}{T_H}\right)dy.
\end{align}
The values of $\langle p_T\rangle$ and $\langle \theta\rangle$
from Eqs. (33)--(36) can be obtained by a numerical method, if
necessary.

The cold source has the capacity to evaporate light fragments
while leaving behind heavier ones due to its low excitation but
large volume. In contrast, because of its high excitation and
small volume, the hot source can only emit light fragments and not
heavy ones. This implies that light fragments are emitted at two
distinct temperatures, whereas heavy fragments are exclusively
emitted at low temperature. In a special scenario where both cold
and hot sources possess identical temperatures, they effectively
merge into a single entity. For multi-fragment emission processes
occurring in $eA$ collisions, it is advisable to initially apply a
single temperature distribution. Should this approach prove
inadequate in the high $p_T$ ($\theta$)  region, one might
consider incorporating contributions from an additional
temperature.

\subsection{Discussion}

It is important to note that there exists no clear boundary
between soft and hard processes. More broadly speaking, in the
present work, regardless of the type of parton involved, processes
characterized by low momentum transfer are generally classified as
soft or first process, while those with high momentum transfer are
categorized as hard or second process; however, both types
represent violent deep inelastic scattering events. If necessary,
experimental measurements can be further divided into results from
additional processes based on momentum transfer---such as
three-processes or four-processes---although this may introduce
more parameters into the analysis. At energies accessible at
current accelerators and colliders, two process classifications
are typically sufficient unless extremely high momentum transfers
are encountered. In certain cases where high momentum transfer
does not occur, one might rely solely on the soft process for data
fitting. It is noteworthy that the terminology employed in this
work regarding soft and hard processes may differ from some
existing literature~\cite{40,40a,40b,40c,40d}, which often refers
to both types simply as hard process. Furthermore, this study
utilizes a methodology grounded in statistical physics that
diverges from quantum chromodynamics (QCD).

At EIC energies (the center-of-mass energy $\sqrt{s}\approx20-140$
GeV for $eN$ collisions~\cite{40dd}), a significant portion of
deep inelastic scattering is dominated by small Bjorken-$x$
physics~\cite{40e,40f,40g,40h}. Specifically speaking,
interactions between electrons and partons become increasingly
relevant under these conditions. Notably at low photon
virtualities and within large nuclei environments---the fraction
of soft process associated with low momentum transfer becomes
predominant; conversely at large photon virtualities---the
contribution from hard process characterized by high momentum
transfer takes precedence~\cite{40i}. Meanwhile, at small-$x$, the
phenomenon of gluon saturation, driven by the rapid increase in
gluon density~\cite{40j}, will enhance the fraction of low
momentum transfer process while naturally reducing the fraction of
high momentum transfer process. The outcome of this comprehensive
interplay indicates that gluon saturation physics plays a critical
role in the physical framework we are investigating, which
pertains to the first process.

As an application of our hybrid model, we summarize here our
previous work~\cite{27}, which analyzed the multiplicity
distributions of charged particles produced in $e^+p$ collisions.
The relevant data were collected using a multipurpose magnetic
detector (ZEUS) operational at the Hadron-Electron Ring
Accelerator (HERA) located at Deutsches Elektronen-Synchrotron
(DESY)~\cite{41}. In ZEUS Collaboration experiments, the beam
energy for $e^+$ is 27.5 GeV and for protons it is 820 GeV,
corresponding to $\sqrt{s} \approx 300$ GeV~\cite{41a,41b}. The
ZEUS data samples are categorized into four groups based on
different selection criteria: (i) in the Breit frame with bins
defined by $2E_B^{\rm cr}$; where the Breit frame is characterized
by conditions ensuring that the momentum of exchanged virtual
bosons is purely spacelike and $E_B^{\rm cr}$ represents available
energy within this region~\cite{41}; (ii) in bins based on
invariant mass ($M_{eff}$) within the Breit frame; (iii) in
hadronic center-of-mass frame with bins determined by $\gamma^*p$
center-of-mass energy ($W$), where $\gamma^*$ denotes virtual
photons exchanged during $e^+p$ collisions; and (iv) again within
hadronic center-of-mass frame but classified according to
$M_{eff}$.

Based on these four types of data samples measured from $e^+p$
collisions at DESY-HERA by ZEUS Collaboration~\cite{41}, our
previous work successfully fitted charged particle multiplicity
distributions using a single-component Erlang
distribution~\cite{27}. Tables 1 and 2 summarize the values of the
parameters $\langle n_{i1}\rangle$ and $m_1$ obtained from these
fits. Notably, there is no contribution from the second process.
It can be observed that $\langle n_{i1}\rangle$ increases with
rising values of $2E_B^{\rm cr}$, $M_{eff}$, and $W$ in both the
Breit frame and hadronic center-of-mass frame, while $m_1$ remains
approximately invariant across the corresponding data samples. In
$eN$ collisions, the number of participating partons is relatively
small; conversely, a significant number of remaining partons act
as spectators. The parameters characterizing multiple particle
production at the EIC could be preliminarily constrained by data
collected at DESY-HERA.

At energies around a few GeV, our recent work~\cite{41c,41d}
employed the single-component Erlang distribution to study squared
momentum transfer spectra for light mesons such as $\pi^0$,
$\pi^+$, $\eta$, and $\rho^0$. These mesons were produced in
$\gamma^*p$ $\rightarrow$ meson+$N$ process during $ep$ collisions
measured at Thomas Jefferson National Accelerator Facility
(Jefferson Laboratory or JLab)~\cite{41e,41f,41g,41h}.
Additionally, we examined squared momentum transfer spectra for
$\eta$ and $\eta_0$, generated in process where $\gamma^*p
\rightarrow \eta(\eta_0)$+$p$ occurred during $ep$ collisions
conducted at Continuous Electron Beam Accelerator Facility
(CEBAF)~\cite{41i}, Daresbury Laboratory electron synchrotron
(NINA)~\cite{41j}, Cambridge Electron Accelerator
(CEA)~\cite{41k}, Stanford Linear Accelerator Center
(SLAC)~\cite{41l}, DESY~\cite{41m}, and Wilson Laboratory
Synchrotron (WLS)~\cite{41n}. Our findings indicate that related
experimental data can be effectively fitted using the hybrid
model. The value of $M_1$ ranges from 3 to 5---lower than that
observed ($m_1=6-11$) at an energy of $\sqrt{s}=300$ GeV. Herein
we estimate that $M_1=m_1$, with average transverse momenta given
by $\langle p_T\rangle \approx (0.4-1.2)$ GeV/$c$, and for this
same event sample we find that $\langle p_{t1}\rangle \approx
(0.4-1.2)\langle n_{i1}\rangle$ GeV/$c$~\cite{41c,41d}. The
parameters derived from low-energy measurements provide valuable
references for future analyses involving $eN$ or $eA$ collisions
at EIC.

\begin{table*}[htb!]\vspace{0.2cm}
\vspace{0.25cm} \justifying\noindent {\small Table 1. Values of
$\langle n_{i1}\rangle$ and $m_1$ for the fit~\cite{27} to data
samples selected by $2E_B^{\rm cr}$ and $M_{eff}$ in the Breit
frame~\cite{41}. \vspace{-0.35cm}
\begin{center}
\begin{tabularx}{0.75\textwidth}{>{\centering\arraybackslash}X>
{\centering\arraybackslash}X>{\centering\arraybackslash}X>{\centering\arraybackslash}X
>{\centering\arraybackslash}X>{\centering\arraybackslash}X}\\
\hline\hline
$2E_B^{\rm cr}$ (GeV) & $\langle n_{i1}\rangle$ & $m_1$ & $M_{eff}$ (GeV) & $\langle n_{i1}\rangle$ & $m_1$\\
\hline
1.5--4  & 0.38 & 6 & 1.5--4 & 0.43 & 9\\
4--8    & 0.54 & 6 & 4--8   & 0.65 & 9\\
8--12   & 0.70 & 6 & 8--12  & 0.90 & 9\\
12--20  & 0.87 & 6 & 12--20 & 1.12 & 9\\
20--30  & 1.04 & 6 &        &      &\\
30--45  & 1.21 & 6 &        &      &\\
45--100 & 1.45 & 6 &        &      &\\
\hline
\end{tabularx}%
\end{center}}
\end{table*}
\vspace{-0.75cm}

\begin{table*}[htb!]\vspace{0.2cm}
\vspace{0.25cm} \justifying\noindent {\small Table 2. Values of
$\langle n_{i1}\rangle$ and $m_1$ for the fit~\cite{27} to data
samples selected by $M_{eff}$ and $W$ in the hadronic
center-of-mass frame~\cite{41}. \vspace{-0.35cm}
\begin{center}
\begin{tabularx}{0.75\textwidth}{>{\centering\arraybackslash}X>
{\centering\arraybackslash}X>{\centering\arraybackslash}X>{\centering\arraybackslash}X
>{\centering\arraybackslash}X>{\centering\arraybackslash}X}\\
\hline\hline
$M_{eff}$ (GeV) & $\langle n_{i1}\rangle$ & $m_1$ & $W$ (GeV) & $\langle n_{i1}\rangle$ & $m_1$\\
\hline
1.5--4 & 0.44 & 10 & 70--100  & 1.68 & 6\\
4--8   & 0.70 &  9 & 100--150 & 1.94 & 6\\
8--12  & 0.80 & 11 & 150--225 & 2.23 & 6\\
12--20 & 1.15 & 10 &  &  & \\
20--30 & 1.47 & 10 &  &  & \\
\hline
\end{tabularx}%
\end{center}}
\end{table*}

The pseudo(rapidity) distributions of final-state particles have
been investigated in our earlier work~\cite{41o}, which confirms
the effectiveness of the two-cylinder model within the hybrid
framework. In $pp$ collisions across an energy range from
$\sqrt{s}=24$ to 63 GeV~\cite{41p}, the charged particle
pseudorapidity distributions indicate that for low multiplicity
events ($n_{ch}=2-4$), the string lengths are
$L_{y_P}=L_{y_T}=3.20-3.40$, corresponding to soft processes,
while for high multiplicity events ($n_{ch}=20-24$), they are
$L_{y_P}=L_{y_T}=2.00-2.80$, indicative of hard processes.
Furthermore, within this energy range, we observe a rapidity gap
$\Delta y=0.8-1.2$ between projectile and target strings in low
multiplicity events, whereas $\Delta y=0$ is noted in high
multiplicity scenarios. These parameters reflect the penetrability
of the collision system during soft processes, characterized by
significant rapidity shifts, and stopping power during hard
processes, marked by minimal rapidity shift.

In $e^+e^-$ annihilations over an energy range from $\sqrt{s}=14$
to 34 GeV~\cite{41q}, charged particle rapidity distributions
reveal that both string lengths satisfy
$L_{y_P}=L_{y_T}=1.72-2.20$, with a gap of $\Delta y=0.4-0.6$. At
an energy level of $\sqrt{s}=29$ GeV, as multiplicities increase,
both $L_{y_P}$ and $L_{y_T}$ decrease while $\Delta y$ increases
correspondingly. Additionally, as particle or quark jet sizes grow
larger, there is a continued decrease in both string lengths;
however, no clear trend emerges for changes in $\Delta y$. Once
again, these parameters elucidate aspects related to collision
system penetrability and stopping power across different
interaction types. Moreover, lighter mass particles and quark jets
tend to achieve higher velocities more readily than heavier
counterparts---resulting in greater rapidity shifts overall. The
findings mentioned here also offer valuable insights relevant to
analyzing $eN$ collisions at EIC experiments, though our earlier
work~\cite{41o} did not specifically address on pseudo(rapidity)
distributions in $ep$ collisions.

One may compare the multi-particle production and multi-fragment
emission in $eA$ collisions. These two processes exhibit
similarities, namely, both possess a two-temperature structure,
and the two-component Erlang distribution can be employed to
describe their multiplicity distributions. The distinction lies in
that the transverse momentum distribution of multiple particles
can also be characterized by a two-component Erlang distribution,
whereas the transverse momentum distribution of multiple fragments
can be described using a two-component Rayleigh distribution when
applying the classical ideal gas model.

We would like to emphasize the differences in the underlying
physics. The two-component distribution for multi-particle
production arises from two types of events: soft excitation
process occurs during violent collisions between electron and
partons, which are primarily sea quarks and/or gluons; hard
scattering process takes place during more intense collisions
between electron and partons, typically valence quarks. In
contrast, the two-component distribution for multi-fragment
emission is present within similar types of events. Specifically,
within an excited nucleus, regions not involved in $eN$ scattering
act as cold sources while those engaged in $eN$ scattering serve
as hot sources.

It is essential to discuss what constitutes leading charged
particles in high-energy collisions. Although there are varying
perspectives on leading charged particles within the community, we
have a specific designation for this article. We contend that in
very forward/backward pseudorapidity regions, the yield of charged
baryons significantly exceeds that of charged mesons. The excess
portion of charged baryons originates from those baryons that
pre-exist in both projectile and target nuclei. Therefore, when
referring to leading charged particles in this article, we
specifically mean leading charged baryons such as protons rather
than mesons (e.g. $\pi^{\pm}$, $K^{\pm}$), leptons (e.g.
$e^{\pm}$), and neutrons.

The radial isotropic flow effect is not excluded from the
transverse momentum distribution due to its very small value or
near-zero contribution in extremely low-energy collisions.
Generally, radial flow occurs in large collision systems, where
the interaction between the projectile and target leads to an
expansion of the collision system. In $eN$ or $eA$ collisions, one
may consider the radial flow effect to be negligible. Furthermore,
transverse anisotropic flows such as elliptic flow exert a minimal
influence that does not necessitate consideration in the
transverse momentum distribution.

In addition to the effects of leading charged particles and
various flows, we can also examine other nuclear effects. Nuclear
clusters, such as $\alpha$-clusters, can impact the multiplicity
$N_F$ distribution of nuclear fragments with a given charge $Z$.
Consequently, this can affect the average value of $\langle
N_F\rangle$. However, it is important to note that nuclear
clusters do not significantly alter the distribution laws for
$p_T$ and $\theta$ of nuclear fragments. The situations regarding
other nuclear effects---such as projectile penetration, target
stopping power, shadowing over subsequent nucleons, and
correlations between pairs of nucleons---are analogous to those
associated with nuclear clusters.

Moreover, we have employed the concept of temperature within
high-energy collisions. For individual events, it is indeed true
that these systems are so small that applying temperature concepts
may be unsatisfactory. Fortunately, high-energy collisions
represent high-throughput experiments; thus statistical
distributions are derived from numerous events. From a grand
canonical ensemble perspective, statistical laws become
recognizable and applicable under temperature considerations. At
minimum, temperature $T$ reflects the average kinetic energy
across multiple particles which can be extracted from $\langle
p_T\rangle$, as discussed in related
references.~\cite{30a,30b,42}.

\section{Summary}

In summary, the application of a hybrid model for $eA$ collisions
at the EIC allows us to predict the bulk properties of
multi-particle production and multi-fragment emission in these
collisions through statistical distribution laws involving two
components. The production of multiple particles occurs via two
distinct processes: soft excitation and hard scattering, which are
categorized as different types of events. In contrast, multiple
fragments are emitted from two localized regions within the
excited nucleus---namely, low and high temperature sources---which
are classified under the same type of event. The proposed physical
framework and mathematical expressions for $eA$ collisions build
upon our previous research.

The multiplicity and transverse momentum distributions of charged
particles produced in $eA$ collisions can be effectively described
by a two-component Erlang distribution derived from a multi-source
thermal model. The number of participants involved in the soft
excitation process is relatively small due to only an electron and
a limited number of partons (primarily sea quarks and/or gluons)
participating in these collisions. Conversely, during hard
scattering process, there are typically only 2 participants: an
electron and one parton (usually a valence quark). Consequently,
each participant's contribution during the soft process is
naturally less significant than that observed in the hard process.

Similarly, the multiplicity distribution of nuclear fragments
emitted in $eA$ collisions can also be characterized by a
two-component Erlang distribution. Both contributors from low- and
high-temperature sources remain limited due to the constrained
volume available within the excited nucleus. Furthermore, both
transverse momentum and angular distributions for nuclear
fragments approximately follow a two-component Rayleigh
distribution; each component aligns with predictions made by
classical ideal gas model. Relativistic forms for transverse
momentum and angular distributions concerning nuclear fragments
will also be accessible for validation through future studies on
$eA$ collisions at EIC.
\\
\\
\\
{\bf Acknowledgments}

The work of the Shanxi Group was supported by the National Natural
Science Foundation of China under Grant No. 12147215, the Shanxi
Provincial Basic Research Program (Natural Science Foundation)
under Grant No. 202103021224036, and the Fund for Shanxi ``1331
Project" Key Subjects Construction. The work of K.K.O. was
supported by the Agency of Innovative Development under the
Ministry of Higher Education, Science and Innovations of the
Republic of Uzbekistan within the fundamental project No.
F3-20200929146 on analysis of open data on heavy-ion collisions at
RHIC and LHC.
\\
\\
{\bf ORCID}
\\
Fu-Hu Liu, https://orcid.org/0000-0002-2261-6899
\\
Khusniddin K. Olimov, https://orcid.org/0000-0002-1879-8458
\\
\\

\end{document}